\documentclass[10pt,aps,twocolumn,showpacs,pra]{revtex4}
\usepackage{epsfig}
\usepackage{amssymb,amsfonts,amsmath,wrapfig,mathrsfs,mathbbol}
\usepackage{graphicx}
\usepackage{graphics,psfrag}
\begin{document}

\title{A Quantum Field Theory Twist to Photon Angular Momentum}


\author{G.F. Calvo, A. Pic\'{o}n and E. Bagan}
\affiliation{Grup de F\'{\i}sica Te\`{o}rica \& IFAE, Universitat Aut\`{o}noma de Barcelona, 08193 Bellaterra (Barcelona), Spain}

\date{\today}

\begin{abstract}
A quantum field theory approach is put forward to generalize the concept of classical spatial light beams carrying orbital angular momentum to the single-photon level. This quantization framework is carried out both in the paraxial and nonparaxial regimes. Upon extension to the optical phase space, closed-form expressions are found for a photon Wigner representation describing transformations on the orbital Poincar\'{e} sphere of unitarily related families of paraxial spatial modes.  
\end{abstract}

\maketitle

\section{INTRODUCTION}
\label{sect:intro}

Photons are among the main carriers of information. This information can be encoded in their energy, linear momentum, and polarization state. In recent years, another degree of freedom for photons has been recognized: their {\em orbital angular momentum} (OAM). In 1992 Allen and coworkers~\cite{Allen92} showed that optical paraxial cylindrical beams having an azimuthal phase dependence of the form $\exp(il\phi)$ carry a discrete OAM of $l\hbar$ units per photon along their propagation direction. This angular momentum produces a mechanical effect (induces a torque) when suitable light patterns interact with matter; it can be transferred from spatial beams containing phase dislocations on their axis (e.g. optical vortices) to suitable trapped microscopic particles in optical tweezers~\cite{Tweezers}. 
\par
At the quantum level, considerable interest has been brought for quantum information processing exploiting single and entangled photons prepared in a superposition of states bearing a well defined OAM~\cite{Mair,Gabi02,Leach02,Vaziri02,Vaziri03,Gabi04,Langford,Gabi05}. Indeed, one of their main distinguishing features is that, at variance with two-level quantum states, or qubits, OAM photon eigenstates involve the more general case of $d$-level ($d\geq2$) quantum states or qudits. Such a generalization to multidimensional states allows to extend quantum coding alphabets without the need to increase the number of entangled photons, providing also a more secure quantum cryptography. Since fewer photons are needed, the multidimensional approach reduces the decoherence associated to many photon entanglement. Moreover, an striking consequence of such a higher dimensional encoding is that violation of local realism for two maximally entangled qudits is stronger than for two maximally entangled qubits, and increases with $d$~\cite{Vaziri02,Kaszlikowski}. For quantum computation applications, OAM photon eigenstates could even enable to optimize certain quantum computing architectures, where a compromise between the number of required qubit and qudit states exists~\cite{Greentree}.
\par
The aim of this paper is to develop a general quantization scheme of field operators in both the nonparaxial and paraxial regimes of light propagation. Within the nonparaxial regime, the obtained operators posses the suitable phase structure that will straightforwardly allow to proceed, at a later stage, to paraxial field operators. In this regime, the field operators can conveniently be expressed in terms of eigenstate modes of the paraxial orbital and spin angular momentum operators. Our approach is further extended to the optical phase space. The paper is organized as follows: Section II reexamines the problem of the separation of the angular momentum of a classical electromagnetic field into orbital and spin angular momentum components and provides a general approach as to whether such a decomposition is physically meaningful. Section III adapts the previous approach into the field quantization formalism. Section IV presents a powerful scheme, based on the Wigner representation, to describe geometric transformations of photons prepared in states bearing OAM. Conclusions of the paper are drawn in Section V.
\par
\section{CLASSICAL APPROACH REVISITED}
\label{sec:CAR}
Energy, linear momentum and angular momentum constitute the key physical quantities that characterize the electromagnetic field configurations. First of all, they are constants of motion. Their conservation can be cast as a continuity equation relating a density and a flux (tensor) density, or current, associated to the conserved quantity. The total free electromagnetic angular momentum ${\bf J}({\bf r}_{0})$ in a given volume $\mathcal{V}$ with respect to a point ${\bf r}_{0}$ is defined by~\cite{Cohen}
\begin{eqnarray}
{\bf J}({\bf r}_{0})& =& \varepsilon_{0}\int_{\mathcal{V}} d^{3}{\bf r}\left( {\bf r}-{\bf r}_{0}\right)\times\left( {\bf E}\times{\bf B}\right)\nonumber \\
& =& {\bf J}(0) - {\bf r}_{0}\times{\bf P} \; ,
\label{eq:JCAR}
\end{eqnarray}
where ${\bf r}$ is the position vector, ${\bf E}$ and ${\bf B}$ are the electric and magnetic fields, and ${\bf P}$ denotes the total linear momentum of the free electromagnetic field. We shall focus our study on ${\bf J}(0)\equiv{\bf J}$. Notice that ${\bf J}$ is defined in an analogous manner as the angular momentum of a system of massive particles. It can be expressed as the integral of an angular momentum density which is equal to the cross product of ${\bf r}$ with the linear momentum density $\varepsilon_{0}( {\bf E}\times{\bf B})$. The conservation of ${\bf J}$ is guaranteed if the flux of angular momentum through the surface $\mathcal{S}$ enclosing $\mathcal{V}$ vanishes (e.g. for fields that decay sufficiently fast when $\mathcal{V}$ becomes large). This is reflected by the integral form of the continuity equation for the $i$-component of the angular momentum~\cite{Barnett}
\begin{eqnarray}
\frac{\partial J_{i}}{\partial t}=-\int_{\mathcal{S}} M_{li}\,d\mathcal{S}_{l} \; ,
\label{eq:ContinuityJCAR}
\end{eqnarray}
where $M_{li}=\epsilon_{ijk}r_{j}\left[\delta_{kl}\left(\varepsilon_{0}E^{2}+\mu_{0}^{-1}B^{2}\right)/2-\varepsilon_{0}E_{k}E_{l}\right.$ $\left.-\mu_{0}^{-1}B_{k}B_{l}\right]$ is the angular momentum flux tensor~\footnote{Here, Roman letter subindices run through the three components $x$, $y$ and $z$, and repeated indices are summed over all these three components. We also employ the usual Kronecker delta $\delta_{ij}$ and the Levi-Civita tensor $\epsilon_{ijk}$, which takes the value $+1$ if $ijk=123$, $231$, $312$, the value $-1$ if $ijk=321$, $213$, $132$, and is zero otherwise.}. 
\par
It is well known that in classical mechanics the total angular momentum of a system of point matter particles can be separated into two contributions. A first one describes the angular momentum associated with the center of mass motion of the whole system, while the second one refers to the relative angular momentum of the constituent particles with respect to the total center of mass. The center of mass contribution is thus dependent on the choice of the reference frame, whereas the relative part is independent. Both quantities obey independent evolution equations. In quantum mechanics, in addition to these two contributions (which one could consider as giving rise to a purely orbital angular momentum), there is an intrinsic or spin angular momentum (independent of the choice of a reference frame) with no classical analog for point particles. Now, the situation for the free electromagnetic field is more subtle. A similar separation of ${\bf J}$ into {\em strict} orbital and spin angular momentum vectors is known to be impossible because no reference frame exist for the photon in which it is at rest~\cite{Cohen,Jauch,Simmons}. It is however feasible to decompose ${\bf J}$ into two observables~\cite{vanEnk94}, whose physical meaning will be provided below, as
\begin{eqnarray}
{\bf J} = {\bf L}+{\bf S}\; , 
\label{eq:JLS}
\end{eqnarray}
with
\begin{eqnarray}
{\bf L}&=& \varepsilon_{0}\sum_{j}\int_{\mathcal{V}}d^{3}{\bf r}\, E^{\perp}_{j}\left({\bf r}\times\nabla\right)A^{\perp}_{j}\; , 
\label{eq:LCAR}\\
{\bf S}&=& \varepsilon_{0}\int_{\mathcal{V}}d^{3}{\bf r}\,{\bf E}_{\perp}\times{\bf A}_{\perp}\; , 
\label{eq:SCAR}
\end{eqnarray}
where the symbol $\perp$ denotes the transverse component of the fields (recall that transverse components of any field ${\bf F}$ satisfy $\nabla\cdot{\bf F}_{\perp}=0$). Since ${\bf L}$ and ${\bf S}$ involve the transverse part of the vector potential ${\bf A}$, they are gauge invariant. Notice that, at variance with ${\bf L}$, ${\bf S}$ is independent of the definition of the origin of the coordinate system. As it occurs for ${\bf J}$, one may show that both ${\bf L}$ and ${\bf S}$ satisfy continuity equations similar to Eq.~(\ref{eq:ContinuityJCAR}).
\par
Within a purely classical description of paraxial light propagation, the total optical angular momentum along the propagation direction can be decomposed into spin and OAM contributions, each associated with polarization and phase distribution of the beam, respectively~\cite{Allen92,Barnett94} (see also the excellent reviews and references therein on this subject~\cite{OAMbook,Santamato}). As we will prove below, these two contributions correspond to the paraxial versions of ${\bf L}$ and ${\bf S}$ along that same direction. Our framework is completely general, thus allowing to envisage the proper quantization of the fields at a later stage.
\par
We begin by considering in the Coulomb gauge the well-known expansion in the continuous plane-wave basis of the vector potential ${\bf A}$ (which is henceforth assumed to be transverse)
\begin{eqnarray}
{\bf A}({\bf r},t) &=& \sum_{\sigma}\int \frac{d^{3}{\bf k}}{\left(16\pi^{3}\varepsilon_{0}c\vert {\bf k}\vert\right)^{1/2}}\nonumber\\
&\times& \left[ {\boldsymbol \epsilon}_{\sigma}({\bf k})\,\alpha_{\sigma}({\bf k})\,e^{i\left({\bf k}\cdot{\bf r}-c\vert{\bf k}\vert t \right)}+\;\textrm{c. c.}\right]\; ,
\label{eq:CARAplanewave}
\end{eqnarray}
where $\alpha_{\sigma}({\bf k})$ are the complex amplitudes corresponding to the two circular polarization unit vectors ${\boldsymbol \epsilon}_{\sigma}({\bf k})$ ($\sigma=+1$ for right-handed and $\sigma=-1$ for left-handed). They satisfy ${\boldsymbol \epsilon}_{\sigma}({\bf k})\cdot{\boldsymbol \epsilon}_{\sigma'}^{*}({\bf k})=\delta_{\sigma\sigma'}$ and ${\bf k}\cdot{\boldsymbol \epsilon}_{\sigma}({\bf k})=0$. Expansion (\ref{eq:CARAplanewave}) is a solution of the d'Alembert wave equation.
\par
Our first aim is to derive an equivalent expression to Eq.~(\ref{eq:CARAplanewave}) which could then lead in a natural way to its corresponding paraxial limit. The essential feature of any paraxial field is that it can be represented as an envelope field modulating a carrier plane wave with wave vector ${\bf k}_{0}$. Without loss of generality we take ${\bf k}_{0}=k_{0}{\bf u}_{z}$, with $k_{0}>0$ and unitary vector ${\bf u}_{z}$, which implies that the carrier plane wave propagates along the positive $z$ direction.
\par
Let us introduce the trivial identity 
\begin{eqnarray}
\int_{0}^{\infty} dk_{0}\frac{e^{ik_{0}(z-ct)}}{e^{ik_{0}(z-ct)}}\,\delta\left[ k_{0}-f({\bf k})\right] = 1
\; ,
\label{eq:trivialidentity}
\end{eqnarray}
where the function $f({\bf k})>0$, which at this stage can be any arbitrary function, will be specified below. Upon multiplication of Eq.~(\ref{eq:CARAplanewave}) by (\ref{eq:trivialidentity}) and rearranging, we have
\begin{eqnarray}
{\bf A}({\bf r},t) = \int_{0}^{\infty} dk_{0}e^{ik_{0}(z-ct)} {\bf A}_{k_{0}}({\bf r},t)
\, .
\label{eq:CARAplanewave2}
\end{eqnarray}
Now, by imposing ${\bf A}_{k_{0}}({\bf r},t)$ to obey the paraxial wave equation, we obtain the explicit dependence of $f({\bf k})$
\begin{eqnarray}
f({\bf k}) = \frac{k_{z}+\sqrt{k_{z}^{2}+2q^{2}}}{2} \; ,
\label{eq:drk0k}
\end{eqnarray}
where we have denoted ${\bf k}={\bf q}+k_{z}{\bf u}_{z}$, ${\bf q}$ and $k_{z}$  being the transverse (in the $x$-$y$ plane) and longitudinal (along the $z$-axis) wave vectors, respectively. Equation~(\ref{eq:drk0k}) yields a dispersion relation between $k_{0}$ and ${\bf k}$ which gives rise to the constraint imposed by the argument $f({\bf k})$ in the delta function of Eq.~(\ref{eq:trivialidentity}). Namely, $\delta\left[ k_{0}-f({\bf k})\right]=\delta\left[ k_{z}-\left(k_{0}-\frac{q^{2}}{2k_{0}}\right)\right]\left( 1+\vartheta^{2}\right)$, with $\vartheta=q/\sqrt{2k_{0}^{2}}$ a parameter that governs the degree of paraxiality. 
\par
The above transformations enable us to map the variable $k_{z}$, defined in the whole real axis, into the more convenient positive variable $k_{0}$. In this way, one may easily carry the integration with respect to $k_{z}$ and cast Eq.~(\ref{eq:CARAplanewave}) in a form that displays both the paraxial and nonparaxial contributions in a more transparent fashion
\begin{widetext}
\begin{eqnarray}
{\bf A}({\bf r},t)& = &\sum_{\sigma}\int_{0}^{\infty} dk_{0}\int d^{2}{\bf q}\left[ \frac{\left(1+\vartheta^{2}\right)^{2}}{16\pi^{3}\varepsilon_{0}ck_{0}\sqrt{1+\vartheta^{4}}}\right]^{1/2}\Bigl\{ {\boldsymbol \epsilon}_{\sigma}\left[{\bf q},k_{0}(1-\vartheta^{2})\right] \alpha_{\sigma}\left[{\bf q},k_{0}(1-\vartheta^{2})\right]\Bigr. \nonumber\\
&\times& \left. e^{ik_{0}(z-ct)}\exp \left[i{\bf q}\cdot{\bf r}_{\perp}-ik_{0}\vartheta^{2}z -ick_{0}\left(\sqrt{1+\vartheta^{4}}-1\right)t\right] + \;\textrm{c. c.}\right\} \, .
\label{eq:CARAplanewave3}
\end{eqnarray}
\end{widetext}
Isotropy of free space allows us to choose the unit polarization vectors ${\boldsymbol \epsilon}_{\sigma}\left[{\bf q},k_{0}(1-\vartheta^{2})\right]$ in a variety of ways. In what follows, it will be convenient to write them as
\begin{eqnarray}
{\boldsymbol \epsilon}_{\sigma}\left[{\bf q},k_{0}(1-\vartheta^{2})\right] & = & \frac{e^{-i\sigma\varphi}}{\sqrt{2}}\Biggl[ {\bf u}_{\rho}\frac{\left(1-\vartheta^{2}\right)}{\sqrt{1+\vartheta^{4}}}\Biggr. \nonumber \\
&-&\Biggl. i\sigma {\bf u}_{\varphi}-{\bf u}_{z}\sqrt{\frac{2\vartheta^{2}}{1+\vartheta^{4}}} \Biggr]\, ,
\label{eq:epsilonsigma}
\end{eqnarray}
where ${\bf u}_{\rho}={\bf q}/q$, ${\bf u}_{\varphi}={\bf u}_{z}\times{\bf q}/q$, and $\varphi$ is the polar angle in cylindrical coordinates. The circular polarization vectors ${\boldsymbol \epsilon}_{\sigma}\left[{\bf q},k_{0}(1-\vartheta^{2})\right]$ are both of them orthogonal to ${\bf q}+ {\bf u}_{z}k_{0}(1-\vartheta^{2})$, as they should. It is important to emphasize that the field (\ref{eq:CARAplanewave3}) is exactly equal to the starting expansion~(\ref{eq:CARAplanewave}). It still obeys the d'Alembert equation for all times $t>0$. However, it now exhibits the suitable structure to perform the paraxial approximation for which $\vartheta\ll 1$. In this limit, Eq.~(\ref{eq:CARAplanewave3}) reduces to
\begin{eqnarray}
{\bf A}_{P}({\bf r},t)&=& \sum_{\sigma}\int_{0}^{\infty}\frac{dk_{0}}{(16\pi^{3}\varepsilon_{0}ck_{0})^{1/2}}\!\int d^{2}{\bf q}\Bigl[ {\boldsymbol\epsilon}_{\sigma}\,\alpha_{\sigma}\left({\bf q},k_{0}\right)\Bigr.\nonumber\\
&\times&\Bigl. e^{ik_{0}(z-ct)} e^{i\left({\bf q}\cdot{\bf r}_{\perp}-k_{0}\vartheta^{2}z\right)} + \;\textrm{c. c.}\,\Bigr] \, ,
\label{eq:CARAparaxial}
\end{eqnarray}
where we have only retained the quadratic dependence on $\vartheta$ in the second phase factor of Eq.~(\ref{eq:CARAplanewave3}) as the relevant paraxial contribution. The polarization vectors ${\boldsymbol \epsilon}_{\sigma}=({\bf u}_{x} - i\sigma {\bf u}_{y})/\sqrt{2}$ are now independent of ${\bf q}$ and $k_{0}$ (${\bf u}_{x}$ and ${\bf u}_{y}$ are the unit vectors along $x$ and $y$ directions). Notice that the structure of the ${\bf q}$-integrand on the right-hand-side of Eq.~(\ref{eq:CARAparaxial}) resembles the well known paraxial angular spectrum~\cite{Goodman}. One may exploit this fact and expand, rather than in transverse plane wave components, into a new transverse state basis: that of Laguerre-Gaussian (LG) modes, $LG_{l,p}({\bf r}_{\perp},z;k_{0})$, and their Fourier-transformed profiles, $\mathcal{LG}_{l,p}({\bf q})$, at $z=0$ (see Appendix for details). Using their closure relation, one has
\begin{eqnarray}
e^{i\left[{\bf q}\cdot{\bf r}_{\perp}-k_{0}\vartheta^{2}z\right]}=\sum_{l,p}\mathcal{LG}^{*}_{l,p}({\bf q})LG_{l,p}({\bf r}_{\perp},z;k_{0}) \; .
\label{eq:CARplanetoLG}
\end{eqnarray}
Since the LG modes constitute a complete, infinite-dimensional basis for the solutions of the paraxial wave equation, any spatial beam satisfying the paraxial wave equation can therefore be represented in the LG basis in terms of an infinite expansion with complex amplitudes $\alpha_{\sigma,l,p}$ defined by 
\begin{eqnarray}
\alpha_{\sigma,l,p}\left(k_{0}\right)=\int d^{2}{\bf q}\,\mathcal{LG}^{*}_{l,p}({\bf q})\,\alpha_{\sigma}\left({\bf q},k_{0}\right)\,  .
\label{eq:CARalphaLG}
\end{eqnarray}
\par
Hence, one may cast Eq.~(\ref{eq:CARAparaxial}) as
\begin{eqnarray}
{\bf A}_{P}({\bf r},t)&=&\sum_{\sigma,l,p}\int_{0}^{\infty}\frac{dk_{0}}{(16\pi^{3}\varepsilon_{0}ck_{0})^{1/2}}\Bigl[ {\boldsymbol \epsilon}_{\sigma}\,\alpha_{\sigma,l,p}\left(k_{0}\right)\Bigr.\nonumber\\
&\times&\Bigl. e^{ik_{0}(z-ct)}LG_{l,p}({\bf r}_{\perp},z;k_{0})+ \;\textrm{c. c.}\,\Bigr] .
\label{eq:CARAparaxial2}
\end{eqnarray}
The paraxial electric and magnetic fields follow from Eq.~(\ref{eq:CARAparaxial2}) via the well-known relations ${\bf E}_{P}=-\partial {\bf A}_{P}/\partial t$ and ${\bf B}_{P}=\nabla\times{\bf A}_{P}$. It is now possible to show, by employing our convenient representation of these paraxial fields into expressions (\ref{eq:LCAR}) and (\ref{eq:SCAR}), that the $z$-components of ${\bf L}$ and ${\bf S}$ are given by
\begin{eqnarray}
L_{z}&=& \sum_{\sigma,l,p} l\int_{0}^{\infty} dk_{0}\, \vert \alpha_{\sigma,l,p}\left(k_{0}\right)\vert^{2}\; , 
\label{eq:CARparaxialL}\\
S_{z}&=& \sum_{\sigma,l,p} \sigma\int_{0}^{\infty} dk_{0}\, \vert \alpha_{\sigma,l,p}\left(k_{0}\right)\vert^{2}\; .
\label{eq:CARparaxialS}
\end{eqnarray}
That is, within the paraxial approximation, the total angular momentum $J_{z}$ along the beam propagation direction, i.e. along $z$, can be decomposed into the so-called {\em orbital} $L_{z}$ and {\em spin} $S_{z}$ angular momenta which are related to the azimuthal phase dependence of the LG mode basis and their corresponding circular polarization state, respectively. Equations (\ref{eq:CARparaxialL}) and (\ref{eq:CARparaxialS}) generalize the well known results of Allen and coworkers~\cite{Allen92} with the remarkable feature that they now posses the appropriate form to carry out the field quantization. Notice also that, since the total energy of any paraxial spatial beam is $H_{P}=\varepsilon_{0}\int d^{3}{\bf r}\left[ E_{P}^{2}+c^{2}B_{P}^{2}\right]/2=\sum_{\sigma,l,p}\int_{0}^{\infty} dk_{0}\, ck_{0}\vert \alpha_{\sigma,l,p}\left(k_{0}\right)\vert^{2}$, in units of $\hbar$, the ratio $L_{z}/H_{P}$ can be conceived from a semiclassical point of view as the OAM per photon.
\par
\section{PARAXIAL QUANTIZATION}
\label{sec:PQ}
In the previous section we have derived general classical paraxial expressions for the orbital and spin angular momenta starting from the continuous plane wave expansions of the fields in free space. In this section we shall undertake the paraxial quantization of the fields. This problem has been considered by several authors in the past (see Ref.~\cite{Kolobov} and references therein). There, the approach was approximated, often requiring perturbation expansions which did not provide any clear and manageable expressions for the paraxial quantum modes, and, more important, they resorted to the quasi-monochromatic approximation which is unsuitable for describing such nonclassical processes as the propagation of broad-band entangled photons generated by spontaneous parametric down conversion.
\par
The framework developed in the previous section can easily be adapted for field operators. The essential step is to substitute the set of complex amplitudes $\alpha$ by creation and annihilation operators. At a suitable stage we will find more convenient to introduce a different set of creation and annihilation operators: those associated with the LG modes. 
\par
The expansion for the vector potential operator $\hat{\bf A}({\bf r},t)$ in terms of continuous plane-wave mode operators [recall Eq.~(\ref{eq:CARAplanewave})] is 
\begin{eqnarray}
 \hat{\bf A}({\bf r},t) &=& \sum_{\sigma}\int d^{3}{\bf k} \left(\frac{\hbar}{16\pi^{3}\varepsilon_{0}c\vert {\bf k}\vert}\right)^{1/2}\nonumber\\
&\times&\left[ {\boldsymbol \epsilon}_{\sigma}({\bf k})\,\hat{a}_{\sigma}({\bf k})\,e^{i\left({\bf k}\cdot{\bf r}-c\vert{\bf k}\vert t \right)}+\;\textrm{h. c.}\right]\; .
\label{eq:QAplanewave}
\end{eqnarray}
The annihilation and creation operators $\hat{a}_{\sigma}$ and $\hat{a}_{\sigma}^{\dagger}$ satisfy the usual canonical commutation rules $[\hat{a}_{\sigma}({\bf k}), \hat{a}_{\sigma'}^{\dagger}({\bf k}')]=\delta_{\sigma\sigma'}\delta^{(3)}({\bf k}-{\bf k}')$. 
\par
Using again the trivial identity~(\ref{eq:trivialidentity}), we can derive the quantized version of Eq.~(\ref{eq:CARAplanewave3})
\begin{widetext}
\begin{eqnarray}
\hat{\bf A}({\bf r},t)& = &\sum_{\sigma}\int_{0}^{\infty} dk_{0}\int d^{2}{\bf q}\left[ \frac{\hbar\left(1+\vartheta^{2}\right)^{2}}{16\pi^{3}\varepsilon_{0}ck_{0}\sqrt{1+\vartheta^{4}}}\right]^{1/2}\Bigl\{ {\boldsymbol \epsilon}_{\sigma}\left[{\bf q},k_{0}(1-\vartheta^{2})\right] \hat{a}_{\sigma}\left[{\bf q},k_{0}(1-\vartheta^{2})\right]\Bigr. \nonumber\\
&\times& \left. e^{ik_{0}(z-ct)} \exp \left[i{\bf q}\cdot{\bf r}_{\perp}-ik_{0}\vartheta^{2}z -ick_{0}\left(\sqrt{1+\vartheta^{4}}-1\right)t\right] + \;\textrm{h. c.}\right\} \, .
\label{eq:QAplanewave3}
\end{eqnarray}
\end{widetext}
The unit polarization vectors ${\boldsymbol \epsilon}_{\sigma}\left[{\bf q},k_{0}(1-\vartheta^{2})\right]$ are given by Eq.~(\ref{eq:epsilonsigma}). Equation~(\ref{eq:QAplanewave3}), together with its classical counterpart~(\ref{eq:CARAplanewave3}), constitute the first main results of this work. We stress that the field operator (\ref{eq:QAplanewave3}) is exactly equal to expansion~(\ref{eq:QAplanewave}), and thus, it obeys the d'Alembert wave equation for any time $t>0$. 
\par
We may now build the paraxial version of Eq.~(\ref{eq:QAplanewave3}). As mentioned previously, this corresponds to the limit with $\vartheta\ll 1$. Recalling the transformation~(\ref{eq:CARplanetoLG}) for the LG modes basis, Eq.~(\ref{eq:QAplanewave3}) reduces to the form
\begin{eqnarray}
\hat{\bf A}_{P}({\bf r},t)&=& \sum_{\sigma,l,p}\int_{0}^{\infty}dk_{0}\left(\frac{\hbar}{16\pi^{3}\varepsilon_{0}ck_{0}}\right)^{1/2}\Bigl[ {\boldsymbol \epsilon}_{\sigma}\,\hat{a}_{\sigma,l,p}\left(k_{0}\right)\Bigr.\nonumber\\
&\times&\Bigl. e^{ik_{0}(z-ct)}LG_{l,p}({\bf r}_{\perp},z;k_{0})+ \;\textrm{h. c.}\,\Bigr] ,
\label{eq:QAparaxial}
\end{eqnarray}
where the circularly-polarized polarized vectors ${\boldsymbol \epsilon}_{\sigma}= ({\bf u}_{x} - i\sigma {\bf u}_{y})/\sqrt{2}$ are again independent of ${\bf q}$ and $k_{0}$. At this point, we have introduced the LG mode annihilation operators
\begin{eqnarray}
\hat{a}_{\sigma,l,p}\left(k_{0}\right)=\int d^{2}{\bf q}\,\mathcal{LG}^{*}_{l,p}({\bf q})\, \hat{a}_{\sigma}\left({\bf q},k_{0}\right) ,
\label{eq:QLGan}
\end{eqnarray}
which, by employing the commutation relations $[\hat{a}_{\sigma}({\bf q},k_{0}),\hat{a}_{\sigma'}^{\dagger}({\bf q}',k_{0}')]=\delta_{\sigma\sigma'}\delta^{(2)}({\bf q}-{\bf q}')\delta(k_{0}-k_{0}')$, and the orthonormalization conditions for the LG modes, satisfy the commutators
\begin{eqnarray}
\!\!\!\!\!\! \left[\hat{a}_{\sigma,l,p}\left(k_{0}\right), \hat{a}_{\sigma',l',p'}^{\dagger}(k_{0}')\right]=\delta_{\sigma\sigma'}\delta_{ll'}\delta_{pp'}\delta(k_{0}-k_{0}')\, .
\label{eq:acommutators}
\end{eqnarray}
\par
We have now at our disposal all the necessary ingredients to examine the quantization of angular momentum. For massless particles, such as the photon, the only physically meaningful component of the total angular momentum operator $\hat{\bf J}$ is the one along their propagation direction~\cite{Jauch}. Of course, one may formally quantize the decomposition of $\hat{\bf J}=\hat{\bf L}+\hat{\bf S}$ where the operators $\hat{\bf L}$ and $\hat{\bf S}$ are given by Eqs.~(\ref{eq:LCAR}) and (\ref{eq:SCAR}) with the classical fields replaced by field operators. They can be cast in the form~\cite{Arnaut}
\begin{eqnarray}
\hat{\bf L}&=& -\frac{i\hbar}{2}\sum_{j,\sigma,\sigma'}\int d^{3}{\bf k}\left\{{\epsilon}_{j,\sigma}^{*}({\bf k})\,\hat{a}_{\sigma}^{\dagger}({\bf k})e^{ic\vert{\bf k}\vert t}\left[ {\bf k}\times\nabla_{\bf k}\right]\right. \nonumber\\
&\times&\left. {\epsilon}_{j,\sigma'}({\bf k})\,\hat{a}_{\sigma'}({\bf k})e^{-ic\vert{\bf k}\vert t}- \;\textrm{h. c.}\right\} , 
\label{eq:QL}
\end{eqnarray}
and
\begin{eqnarray}
\hat{\bf S}= \hbar \sum_{\sigma}\sigma\int d^{3}{\bf k}\, \frac{\bf k}{k}\,\hat{a}_{\sigma}^{\dagger}({\bf k})\hat{a}_{\sigma}({\bf k})\; . 
\label{eq:QS}
\end{eqnarray}
Notice, however, that neither component of $\hat{\bf L}$ and $\hat{\bf S}$ is a true angular momentum operator since they do not fulfill the usual SU(2) commutation relations. Instead, they read as: $[\hat{S}_{i},\hat{S}_{j}]=0$, $[\hat{L}_{i},\hat{S}_{j}]=i\hbar\varepsilon_{ijk}\hat{S}_{k}$, and $[\hat{L}_{i},\hat{L}_{j}]=i\hbar\varepsilon_{ijk}(\hat{L}_{k}-\hat{S}_{k})$. Nevertheless, it is still possible to choose any two commuting components $\hat{L}_{i}$ and $\hat{S}_{i}$, and thus construct simultaneous eigenstates of these operators. 
\par
Within the paraxial approximation discussed above, we may concentrate ourselves on the two commuting operators $\hat{L}_{z}$ and $\hat{S}_{z}$. Upon inverting Eq.~(\ref{eq:QLGan}), yielding $\hat{a}_{\sigma}\left({\bf q},k_{0}\right) = \sum_{l,p}\mathcal{LG}_{l,p}({\bf q})\,\hat{a}_{\sigma,l,p}\left(k_{0}\right)$, we find from Eqs.~(\ref{eq:QL}) and (\ref{eq:QS}) that the paraxial OAM and spin operators are finally given by
\begin{eqnarray}
\hat{L}_{z}&=& \hbar\sum_{\sigma,l,p}l\int_{0}^{\infty} dk_{0}\,\hat{a}_{\sigma,l,p}^{\dagger}\left(k_{0}\right)\hat{a}_{\sigma,l,p}\left(k_{0}\right)\; , 
\label{eq:QLparaxial}\\
\hat{S}_{z}&=& \hbar \sum_{\sigma,l,p}\sigma\int_{0}^{\infty} dk_{0}\,\hat{a}_{\sigma,l,p}^{\dagger}\left(k_{0}\right)\hat{a}_{\sigma,l,p}\left(k_{0}\right)\; . 
\label{eq:QSparaxial}
\end{eqnarray}
\par
This means that the most general paraxial one-photon state can then be described as consisting of arbitrary superpositions of eigenstates of $\hat{L}_{z}$ and $\hat{S}_{z}$ 
\begin{eqnarray}
\vert \psi \rangle = \sum_{\sigma,l,p}\int_{0}^{\infty} dk_{0}\, C_{\sigma,l,p}(k_{0})\,\hat{a}_{\sigma,l,p}^{\dagger}\left(k_{0}\right)\vert 0 \rangle\; , 
\label{eq:onephotonparaxial}
\end{eqnarray}
where $\vert 0 \rangle$ is the vacuum state. The complex coefficients $C_{\sigma,l,p}(k_{0})$ satisfy the normalization condition $\sum_{\sigma,l,p}\int_{0}^{\infty} dk_{0}\vert C_{\sigma,l,p}(k_{0})\vert^{2}=1$, and can be interpreted as the probability amplitudes for finding the photon in an eigenstate $\vert \sigma,l,p,k_{0}\rangle$ (Fock state) with circular polarization $\sigma$, wave vector $k_{0}$ along the $z$-axis and corresponding to a LG mode having indices $l$ and $p$, that is, with a well defined spin and OAM in the direction of the $z$-axis.
\par
Interestingly enough, if one uses Eq.~(\ref{eq:QLGan}) and rewrites the one-photon state (\ref{eq:onephotonparaxial}) as
\begin{eqnarray}
\vert \psi \rangle =\sum_{\sigma,l,p}\int_{0}^{\infty}\!dk_{0}\!\int d^{2}{\bf q}\,f_{\sigma,l,p}({\bf q},k_{0})\,\hat{a}_{\sigma}^{\dagger}\left({\bf q},k_{0}\right)\vert 0 \rangle  , 
\label{eq:onephotonparaxial2}
\end{eqnarray}
where $f_{\sigma,l,p}({\bf q},k_{0})=C_{\sigma,l,p}(k_{0})\mathcal{LG}_{l,p}({\bf q})$, one can conceive $f_{\sigma,l,p}({\bf q},k_{0})$ as the components of the paraxial photon wave function in momentum representation. This implies that, within the paraxial approximation, the energy density of a single photon can be localized in the transverse plane orthogonal to their main propagation direction ($z$-axis) with a Gaussian-dependence falloff. Such a spatial localization is not in contradiction with the exponential (but less than Gaussian) falloff localization limit shown by Bialynicki-Birula~\cite{Iwo} which applies to three (and thus to nonparaxial photons) rather than to two spatial dimensions (see also Refs.~\cite{Eberly,Walmsley,Saari,Hawton} for closely related discussion on this). In fact, the Paley-Wiener theorem still affects the maximum localization of the energy density along the propagation direction. Moreover, the characteristic length that controls the transverse spatial extension of the LG mode cannot certainly take arbitrarily small values, but those which are compatible with the paraxial approximation, i.e. much larger than the photon wavelength $\lambda=2\pi/k_{0}$.
\par
It is important to point out that although the described paraxial quantization has been carried out for LG modes, which constitute the natural basis for operator $\hat{L}_{z}$, it is straightforward to show that one could have chosen as well other paraxial modes such as the Hermite-Gaussian (in Cartesian coordinates) and the Ince-Gaussian~\cite{Bandres} (in elliptical coordinates). Moreover, our nonparaxial expression~(\ref{eq:QAplanewave3}) could readily accommodate photon states in Bessel (diffraction-free) modes~\cite{Durnin,Jauregui} and oblate spheroidal mode solutions of the Helmholtz equation~\cite{Gustavo}, to name just a few examples.
\par
\section{Phase Space Picture of OAM photon states}

Phase space, which is a fundamental concept in classical mechanics, remains useful when passing to quantum mechanics. In a similar fashion with probability density distribution functions in classical systems governed by Liouville dynamics, quasiprobability distributions have been introduced in quantum mechanics. They can provide a description of quantum systems at the level of density operators (although not at the level of state vectors). Among them, the Wigner function stands out because it is real, nonsingular, yields correct quantum mechanical operator averages in terms of phase space integrals and possess positive-definite marginal distributions~\cite{Hillery,Lee,Schleich}. It is, however, only positive for Gaussian pure states, according to the Hudson-Piquet theorem.
\par
By exploiting the analogy between classical and quantum mechanics with geometrical and wave optics, Wigner distributions have been developed in the context of classical wave optics of both coherent and partially coherent light fields~\cite{Bastiaans79,Sundar,Dragoman}, where they are Fourier-related to the cross-spectral densities. Particularly outstanding has been the symplectic invariant approach by Simon and Mukunda~\cite{Simon00a}, which has been applied to anisotropic Gaussian Schell-model beams via the relation between ray-transfer matrices of first-order optical systems and unitary (metaplectic) operators acting on wave amplitudes and cross-spectral densities.
\par
It is well known that a convenient way to visualize the transformation of qubits is provided by the Poincar\'{e} (also known as the Bloch) sphere representation. For polarization states, the north and south poles correspond to right- $\vert \sigma = + 1\rangle$ and left-handed $\vert \sigma = -1 \rangle$ circularly polarized eigenstates, respectively. More generally, any completely polarized state can be described as a linear superposition of left- and right-handed circular polarization in the form (up to a global phase)
\begin{eqnarray}
\vert \theta, \varphi\rangle = \cos\frac{\theta}{2}\,\vert\sigma = +1 \rangle + e^{i\varphi}\sin\frac{\theta}{2}\,\vert\sigma = -1 \rangle \, , \label{eq:spinstates}
\end{eqnarray}
which, on the Poincar\'{e} sphere, corresponds to a point on the surface having angular coordinates $\theta$ and $\varphi$. 
\par
In an analogous manner with the above picture for polarization states, a Poincar\'{e} sphere was introduced by Padgett and Courtial to represent paraxial first-order-mode spatial beams carrying OAM~\cite{Padgett99}. Its underlying SU(2) symmetry was subsequently shown~\cite{Agarwal99}. In this picture, the poles of the sphere correspond to LG modes with radial-node number $p=0$ and topological charge $l=\pm 1$ (plus and minus standing for the north and south poles, respectively). Hence, in complete analogy with Eq.~(\ref{eq:spinstates}), any state on the first-order-mode sphere can be written as
\begin{eqnarray}
\vert \theta, \varphi\rangle_{l=1,p=0} & = & \cos\frac{\theta}{2}\,\vert l=1,p=0\rangle \nonumber\\
&+& e^{i\varphi}\sin\frac{\theta}{2}\,\vert l=-1,p=0\rangle \, . \label{eq:fostates}
\end{eqnarray}
This superposition generates other structurally stable states. For example, in the equatorial plane of the sphere, combinations with $\theta=\pi/2$ and $\varphi=0$ ($\theta=\pi/2$, $\varphi=\pi$) give rise to Hermite-Gaussian modes $\vert n_{x},n_{y}\rangle$ having indexes $n_{x}=1$, $n_{y}=0$ ($n_{x}=0$, $n_{y}=1$). 
\par
Though the first-order orbital Poincar\'{e} sphere constitutes an elegant framework to represent families of states bearing OAM and their transformation, as points and paths connecting these points on the sphere, higher-order modes cannot be described by Eq.~(\ref{eq:fostates}). That is, states on higher-order Poincar\'{e} spheres involve more complex superpositions of LG mode states. A generalization would provide, among the many possibilities, direct means to visualize the development of geometric phases in optical beams~\cite{VanEnk,Galvez} and photon states. 
\par
\begin{figure}
\begin{center}
\hspace*{-0.0cm}
\hbox{\vbox{\vskip -0.0cm \epsfig{figure=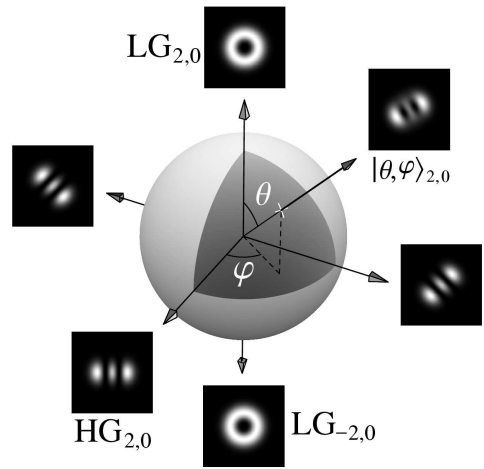, scale=0.7}}}
\end{center}
\vspace*{-0.4cm}
\caption{Orbital Poincar\'{e} sphere of second-order-modes. The poles of the sphere correspond to Laguerre-Gaussian modes with $l =\pm 2$ and $p=0$ (positive and negative signs for north and south poles, respectively). The states in the equatorial plane ($\theta = \pi/2$) with $\varphi = 0$, and $\varphi = \pi$ yield the Hermite-Gaussian modes having indexes $n_{x}=2$, $n_{y}=0$, and $n_{x}=0$, $n_{y}=2$ (not shown), respectively.}
\label{fig:Poincare}
\end{figure}

Recently, we were able to carry out the abovementioned generalization to all higher-order orbital Poincar\'{e} spheres, and extend the concept of geometric phases in paraxial optical beams under continuous mode transformations~\cite{Calvo}. Via an SU(2) Lie-group operator algebra, we mapped spin coherent states ~\cite{Barnettbook} into families of spatial modes carrying OAM and belonging to such generalized Poincar\'{e} spheres. Remarkably, these families of spatial modes could be represented in a compact form by resorting to the Wigner function formalism, allowing to reveal their hidden symmetries. Our aim here is to explicitly show this generalization by constructing sets of OAM photon states $\vert \theta, \varphi\rangle_{l,p}$ as points ($\theta,\varphi$) on orbital Poincar\'{e} spheres $\mathcal{O}_{l,p}$ (which can be labeled by $l$ and $p$ or equivalently by $l$ and the sphere-order $N= 2p + \vert l\vert\geq 0$). The continuum of states on $\mathcal{O}_{l,p}$ can be generated from LG mode states~\footnote{For simplicity, we assume that all states $\vert l,p\rangle$ have equal polarization $\sigma$ and propagate with wave vector $k_{0}$ along $z$. Also, without loss of generality, we consider all $\mathcal{O}_{l,p}$ having $l\geq 0$. Spheres with $l<0$ exhibit a state configuration identical to those with $l>0$, after inversion with respect to their centers.} $\vert l, p\rangle=\hat{a}_{l,p}^{\dagger}\vert 0 \rangle$ (fixed on the poles of the sphere) through the following unitary operations:
\begin{eqnarray}
\vert \theta, \varphi\rangle_{l,p} =\exp(-i\theta\,\hat{\boldsymbol{\mathcal{L}}}\cdot{\bf u}_{\varphi})\vert l, p\rangle\equiv \hat{U}(\theta, \varphi)\vert l, p\rangle . \label{eq:OAMstates}
\end{eqnarray}
Here, $\hat{\boldsymbol{\mathcal{L}}}$ is a true angular momentum operator (not to be confused with $\hat{\bf L}$, see below) and ${\bf u}_{\varphi}=(-\sin\varphi,\cos\varphi,0)$ a unit vector in the equatorial plane of the sphere. The action of the unitary (metaplectic) operator $\hat{U}$ on states $\vert l, p\rangle$ can be interpreted as a counterclockwise rotation of $\theta$ about ${\bf u}_{\varphi}$ that takes the axis containing the poles into the direction with unit vector ${\bf u}_{r}=(\cos\varphi\sin\theta,\sin\varphi\sin\theta,\cos\theta)$. The important observation is that such a rotation gives rise to multidimensional superpositions of LG mode states that can be cast as
\begin{eqnarray}
\vert \theta, \varphi\rangle_{l,p} = \sum_{l'=-l}^{l}\sum_{p'=0}^{p}C_{l',p'}(\theta, \varphi;l,p)\,\vert l',p'\rangle\; ,
\label{eq:OAMPoincarephoton}
\end{eqnarray}
where $2p'+\vert l'\vert = 2p + \vert l\vert \equiv N$. The complex coefficients $C_{l',p'}(\theta, \varphi;l,p)$ depend on the point $(\theta,\varphi)$ on the $N$th-order Poincar\'{e} sphere $\mathcal{O}_{l,p}$. One may easily verify that Eq.~(\ref{eq:OAMPoincarephoton}) includes the particular case of Eq.~(\ref{eq:fostates}) with $C_{1,0}(\theta, \varphi;1,0)=\cos\theta/2$, $C_{0,0}(\theta, \varphi;1,0)=0$ and $C_{-1,0}(\theta, \varphi;1,0)=e^{i\varphi}\sin\theta/2$. Figure \ref{fig:Poincare} depicts representative modes associated to their angular orientations on the second-order-mode $\mathcal{O}_{l,p}$ sphere ($N=l =2$).
\par
\par
The ($\theta,\varphi$)-distribution of states~(\ref{eq:OAMstates}) on $\mathcal{O}_{l,p}$ can be obtained by employing the Wigner function representation. In the optical phase space, let ${\bf r}_{\perp}=(x,y)$ and ${\bf p}=(p_{x},p_{y})$ denote the transverse position and momentum (normalized wave vector) variables, respectively, and $\hat{\bf r}_{\perp}$, $\hat{\bf p}$ be the associated canonical Hermitian operators. The only nonvanishing commutation relations among these operators are $\left[ \hat{x},\hat{p}_{x}\right]=\left[ \hat{y},\hat{p}_{y}\right]=i\lambdabar$. The reduced wavelength $\lambdabar=\lambda/(2\pi)=1/k_{0}$ plays here the optical analog of $\hbar$. For convenience, we arrange the phase space variables and the canonical operators in column vectors ${\boldsymbol \zeta}=(x,y,p_{x},p_{y})$, and $\hat{\boldsymbol\zeta}=(\hat {x},\hat{y},\hat{p}_{x},\hat{p}_{y})$. In terms of $\hat{\boldsymbol\zeta}$, the components of the operator $\hat{\boldsymbol{\mathcal{L}}}$ are 
\begin{subequations}
\label{eq:Ls}
\begin{eqnarray}
\hat{\mathcal{L}}_{x} & = & \frac{\hat{x}^{2}-\hat{y}^{2}}{2w_{0}^{2}}+\frac{(\hat{p}_{x}^{2}-\hat{p}_{y}^{2})w_{0}^{2}}{8\lambdabar^{2}} , \label{eq:Lsx}\\
\hat{\mathcal{L}}_{y} & = & \frac{\hat{x}\hat{y}}{w_{0}^{2}}+\frac{\hat{p}_{x}\hat{p}_{y}w_{0}^{2}}{4\lambdabar^{2}} , \label{eq:Lsy}\\
\hat{\mathcal{L}}_{z} & = & \frac{\hat{x}\hat{p}_{y}-\hat{y}\hat{p}_{x}}{2\lambdabar} . \label{eq:Lsz}
\end{eqnarray}
\end{subequations}
They satisfy the usual SU(2) angular momentum commutators $[\hat{\mathcal{L}}_{i},\hat{\mathcal{L}}_{j}]=i\epsilon_{ijk}\hat{\mathcal{L}}_{k}$ (at variance with the components of operator $\hat{\bf L}$). Of these SU(2) generators, only $\hat{\mathcal{L}}_{z}$ represents real spatial rotations on the transverse $x$-$y$ plane, whereas $\hat{\mathcal{L}}_{x}$ and $\hat{\mathcal{L}}_{y}$ represent simultaneous rotations in the four-dimensional phase-space: $\hat{\mathcal{L}}_{x}$ produces rotations in the $x$-$p_{x}$ and $y$-$p_{y}$ planes by equal and opposite amounts, whereas $\hat{\mathcal{L}}_{y}$ gives rise to rotations in the $x$-$p_{y}$ and $y$-$p_{x}$ planes by equal amounts. Hence, only $\hat{\mathcal{L}}_{z}$ does actually correspond (it is proportional) to component $\hat{L}_{z}$ of Eq.~(\ref{eq:QLparaxial}). There is, however, no analog correspondence between $\hat{\mathcal{L}}_{x}$ and $\hat{\mathcal{L}}_{y}$ and the $x$ and $y$ components of the operator $\hat{\bf L}$, respectively. Operators $\hat{\mathcal{L}}_{x}$ and $\hat{\mathcal{L}}_{y}$ necessarily involve a change of the spatial modes.
\par
The Wigner representation of a photon in a pure state $\psi({\bf p})=\langle {\bf p}\vert \psi \rangle$ is
\begin{eqnarray}
W({\boldsymbol\zeta})&=&\frac{1}{(2\pi\lambdabar)^{2}}\int_{-\infty}^{\infty}\textrm{d}^{2}{\boldsymbol\xi}\,\exp{(i{\bf r}_{\perp}\cdot{\boldsymbol\xi}/\lambdabar)}\nonumber\\
&\times&\psi({\bf p}+\frac{1}{2}{\boldsymbol\xi})\,\psi^{*}({\bf p}-\frac{1}{2}{\boldsymbol\xi}) . \label{eq:Wigner}
\end{eqnarray}
\par
As shown recently in Ref.~\onlinecite{Calvo}, it is possible to obtain the Wigner representation of states $\vert \theta, \varphi\rangle_{l,p}$ without explicitly calculating the integrals in Eq.~(\ref{eq:Wigner}). To this end, we invoke two remarkable properties: (i) On account of the Stone-von Neumann theorem, unitary operators whose generators are quadratic in $\hat{\boldsymbol\zeta}$ [such as $\hat{U}(\theta,\varphi)$] induce linear canonical transformations, $T:\hat{\boldsymbol\zeta}'\rightarrow T\hat{\boldsymbol\zeta}$, in the optical phase space; (ii) under the action of $T$ the Wigner function experiences a simple point transformation $W({\boldsymbol\zeta})\rightarrow W'({\boldsymbol\zeta})=W(T^{-1}{\boldsymbol\zeta})$~\cite{Simon00a}. In our case, the linear canonical transformation generated by the quadratic operators~(\ref{eq:Ls}) results from the relation $T{\hat {\boldsymbol\zeta}}=\hat{U}^{-1}\hat {\boldsymbol \zeta}\hat{U}$, and reads as
\begin{eqnarray}
T=\left(\!\begin{array}{cccc}
c_{\theta}&0&-z_{0}\, s_{\theta}\,s_{\varphi}&z_{0}\, s_{\theta}\,c_{\varphi} \\
0&c_{\theta}&z_{0}\, s_{\theta}\,c_{\varphi}&z_{0}\, s_{\theta}\,s_{\varphi} \\
\frac{s_{\theta}\,s_{\varphi}}{z_{0}}&-\frac{s_{\theta}\,c_{\varphi}}{z_{0}}&c_{\theta}&0\\
-\frac{s_{\theta}\,c_{\varphi}}{z_{0}}&-\frac{s_{\theta}\,s_{\varphi}}{z_{0}}&0&c_{\theta}\end{array}\!\right) ,\label{eq:T}
\end{eqnarray}
where $c_{\theta}=\cos(\theta/2)$, $s_{\theta}=\sin(\theta/2)$, $c_{\varphi}=\cos\varphi$, $s_{\varphi}=\sin\varphi$ (recall that $z_{0} = w_{0}^{2}/(2\lambdabar)$ is the Rayleigh range and $w_{0}$ the mode width at $z=0$). Notice that $T$ has the form of a symplectic ray-transfer matrix of a generally anisotropic first-order system~\cite{KurtBernardo,Alieva05}; it is an element of the symplectic group Sp$(4,R)$, that is, det$\;T=1$, and, under transposition, $T\Lambda T^{t}=T^{t}\Lambda T=\Lambda$, where $\Lambda$ is a real antisymmetric nonsingular 4-dimensional symplectic metric matrix
\begin{eqnarray}
\Lambda=\left(\begin{array}{cc}
0_{2\times 2}&1_{2\times 2}\\
-1_{2\times 2}&0_{2\times 2}\end{array}\right) \, . \label{eq:Lambda}
\end{eqnarray}  
Also, the action of $T$ is independent of the chosen states at $\theta=\varphi=0$, that is, besides  $\vert l, p\rangle$, one could have chosen other state vectors, for instance, Hermite-Gaussian states $\vert n_{x},n_{y}\rangle$ with sphere-order $N=n_{x}+n_{y}=2p+\vert l\vert$.
\par
The key point is thus to observe that owing to the unitary relation (\ref{eq:OAMstates}) between states belonging to the same sphere $\mathcal{O}_{l,p}$, knowledge of the Wigner function of {\em any} given state on $\mathcal{O}_{l,p}$ allows one to determine the Wigner function of all states on that same sphere. LG states constitute the convenient choice here. Using their Wigner representation~\cite{Simon00b}, together with property (ii) and Eq.~(\ref{eq:T}), the found normalized Wigner function is~\cite{Calvo}
\begin{eqnarray}
W_{l,p}({\boldsymbol\zeta};\theta,\varphi) &=& \frac{(-1)^{N}}{\pi^{2}\lambdabar^{2}}e^{-Q_{0}}L_{\frac{N-l}{2}}\left(Q_{0}-4{\bf Q}\cdot{\bf u}_{r}\right)\nonumber\\
&\times& L_{\frac{N+l}{2}}\left(Q_{0}+4{\bf Q}\cdot{\bf u}_{r}\right) , \label{eq:WignerOAM}
\end{eqnarray}
where $Q_{0} =2[x^{2}+y^{2} + (p_{x}^{2}+p_{y}^{2})z_{0}^{2}]/w^{2}$, $L_{m}(\eta)$ are the $m$th order Laguerre polynomials, and the quadratic polynomials ${\bf Q}({\boldsymbol\zeta})\equiv(Q_{x},Q_{y},Q_{z})$ follow from $\hat{\mathcal{L}}_{x}$, $\hat{\mathcal{L}}_{y}$, and $\hat{\mathcal{L}}_{z}$ in Eqs.~(\ref{eq:Ls}) by replacing $\hat{\boldsymbol\zeta}\rightarrow{\boldsymbol\zeta}$. When $\theta =0$ ($\theta =\pi$) one recovers from Eq.~(\ref{eq:WignerOAM}) the Wigner function of LG states $\vert l, p\rangle$ ($\vert -l, p\rangle$). If $\theta =\pi/2$ and $\varphi=0$ ($\theta =\pi/2$ and $\varphi=\pi$) one obtains the Wigner function of Hermite-Gaussian states $\vert n_{x},N-n_{x}\rangle$ ($\vert N-n_{y},n_{y}\rangle$). 
\par
Equation~(\ref{eq:WignerOAM}) is a strictly positive and angle-independent Gaussian distribution only when $l=p=0$ (in this case its associated Poincar\'{e} sphere becomes degenerated, i.e. all points $(\theta,\varphi)$ on the sphere represent the same Gaussian mode state). Moreover, though $W_{l,p}({\boldsymbol\zeta};\theta,\varphi)$ does not explicitly contain the propagation variable $z$, its spatial evolution along $z$ can be fully described by applying a {\em Galilean} boost ${\bf r}_{\perp}\rightarrow {\bf r}_{\perp} - z{\bf p}$.
\par
The orthogonality relations (scalar product) satisfied by states $\vert\theta,\varphi\rangle_{l,p}$ and $\vert\theta',\varphi'\rangle_{l',p'}$ are given by the overlap integral of their associated Wigner functions
\begin{eqnarray}
&&\!\!\!\!\!\!\!\!\!\!\!\!\!\!\!\!\!\!\vert_{l',p'}\langle\theta',\varphi'\vert\theta,\varphi\rangle_{l,p}\vert^{2}\nonumber\\
&=&(2\pi\lambdabar)^{2}\int_{-\infty}^{\infty} \textrm{d}^{4}{\boldsymbol\zeta}\, W_{l',p'}({\boldsymbol\zeta};\theta',\varphi')\, W_{l,p}({\boldsymbol\zeta};\theta,\varphi)\nonumber\\
&=&\left[\sum_{k=0}^{(N-l)/2}\!(-1)^{k}\binom{\frac{N+l}{2}}{k}\binom{\frac{N-l}{2}}{k}\!\left(\frac{1-\tau}{\tau}\right)^{k}\right]^{2}\nonumber\\
&\times&\tau^{N}\delta_{l,l'}\delta_{p,p'}, 
\label{eq:Wignerorthogonality}
\end{eqnarray}
where parameter $\tau=\cos^{2}\left[(\theta-\theta')/2 \right]\cos^{2}\left[(\varphi-\varphi')/2 \right]+\cos^{2}\left[(\theta+\theta')/2 \right]\sin^{2}\left[(\varphi-\varphi')/2 \right]$ (notice that $0\leq\tau\leq 1$). Equation~(\ref{eq:Wignerorthogonality}) implies the following: (i) any two states belonging to different spheres are mutually orthogonal; (ii) if $l>0$ and $p=0$, only states corresponding to antipodal points are mutually orthogonal. However, if $l>0$ and $p>0$, additional points exist on the sphere (apart from the antipodal) where their associated states are also orthogonal [e.g. if $p=1$, Eq.~(\ref{eq:Wignerorthogonality}) vanishes when $\tau=(l+1)/(l+2)$]; (iii) when $l=0$, antipodal points no longer correspond to orthogonal states but to identical states. 
\par
The expectation value $_{l,p}\langle\theta,\varphi\vert\hat{\boldsymbol{\mathcal{L}}}\vert\theta,\varphi\rangle_{l,p}$ may be easily evaluated with the help of the Wigner function~(\ref{eq:WignerOAM}). Since the operators~(\ref{eq:Ls}) do not involve products of noncommuting canonically conjugated operators, their corresponding phase space representation in the Wigner-Weyl ordering is simply given by replacing $\hat{\boldsymbol\zeta}\rightarrow{\boldsymbol\zeta}$ in Eqs.~(\ref{eq:Ls}). Therefore,
\begin{eqnarray}
\langle \hat{\boldsymbol{\mathcal{L}}} \rangle\equiv \,_{l,p}\langle\theta,\varphi\vert\hat{\boldsymbol{\mathcal{L}}}\vert\theta,\varphi\rangle_{l,p} &=&
\int_{-\infty}^{\infty} \textrm{d}^{4}{\boldsymbol\zeta}\, {\boldsymbol{\mathcal{L}}}({\boldsymbol\zeta})W_{l,p}({\boldsymbol\zeta};\theta,\varphi)\nonumber\\
 &=& \frac{l}{2}{\bf u}_{r}\, . 
\label{eq:Lsexpectation}
\end{eqnarray}
Via the Heisenberg-Robertson uncertainty relation and using Eq.~(\ref{eq:Lsexpectation}) we obtain that the variances $\Delta\hat{\mathcal{L}}_{i}$ of operators (\ref{eq:Ls}) satisfy the following inequalities
\begin{eqnarray}
\Delta\mathcal{L}_{i}\,\Delta\mathcal{L}_{j}\geq \frac{1}{2}\vert \epsilon_{ijk}\langle\hat{\mathcal{L}}_{k}\rangle\vert\, . 
\label{eq:Heisenberg}
\end{eqnarray}
In particular, states on the sphere equator (with $\theta=\pi/2$) yield $\Delta\mathcal{L}_{x}\,\Delta\mathcal{L}_{y}\geq 0$. Moreover, the OAM carried by states $\vert\theta,\varphi\rangle_{l,p}$ follows immediately from Eq.~(\ref{eq:Lsexpectation}). The key observation is to notice that the operator $\hat{L}_{z}$, given by Eq.~(\ref{eq:QLparaxial}), for a given polarization $\sigma$ and wave vector $k_{0}$, is related to operator $\hat{\mathcal{L}}_{z}$ by $\hat{L}_{z}=2\hbar\hat{\mathcal{L}}_{z}$. The result, $_{l,p}\langle\theta,\varphi\vert\hat{L}_{z}\vert\theta,\varphi\rangle_{l,p}=l\hbar\cos\theta$, has a very simple geometrical interpretation. It is the projection of the unit vector ${\bf u}_{r}$ corresponding to $\vert \theta, \varphi\rangle_{l,p}$ along the vertical axis of $\mathcal{O}_{l,p}$. Hence, arbitrary states on any sphere bear fractional OAM in units of $\hbar$. The limiting cases, being represented by LG and Hermite-Gaussian states, have the well known $l\hbar$ and zero values, respectively~\cite{Allen92,OAMbook}.
\par
\section{Conclusions}

We have presented a detailed description of classical optical beams and single photon states bearing OAM. Our quantum field theory formalism generalizes previous studies on this subject and, via phase space methods, highlights the inherent symmetries of unitarily related families of paraxial spatial modes, of which, those carrying integer OAM (in units of $\hbar$) constitute one particular subset. The $N$-th order orbital Poincar\'{e} sphere representation enables to visualize mode transformations. These transformations correspond to first-order optical systems with symplectic ray-transfer matrix $T$ [given by Eq.~(\ref{eq:T})] and evidence that it is possible to manipulate single photons prepared in superpositions of OAM states, and, thus, implement single qubit gates (which correspond to particular rotations on the Poincar\'{e} sphere). Combination of these gates with two-photon quantum gates exploiting OAM entangled states generated from parametric down conversion~\cite{Mair,Vaziri02,Vaziri03,Gabi04,Franke,Torres,Walborn04,Gabi05,Law} constitutes a realistic and a fascinating possibility towards quantum computation with linear optical networks~\cite{KLM,OBrien,Gasparoni,Fiorentino,Nemoto}. One could also exploit the multidimensional Hilbert structure of OAM photon states for quantum key distribution (QKD) protocols. At variance with QKD schemes employing polarization states, which only allow transmission of one key bit per photon and require the reference frames of the sender and receiver to be aligned with each other~\cite{Bagan}, OAM photon states are invariant under rotations along their propagation direction. The continuous reference frame alignment monitorization for polarization states may not seem to strong of a restriction for ground-based stations, but it could be an important limitation on a moving station such as a satellite. The distortions created by atmospheric turbulence on OAM photon states could be corrected using two-dimensional filtering techniques that have been proposed for image recovering (adaptive optics) under various degradation mechanisms~\cite{Sadhar}. To conclude, we also wish to point out that detection of phenomena related to highly energetic photons~\cite{Peele} ($X$-rays and Gamma-rays) carrying OAM could be of interest for astrophysical measurements of distant cosmic entities (pulsars, quasars, black holes) through Compton scattering experiments.
\par
\acknowledgments
We gratefully acknowledge financial support from Spanish Ministry of Science and Technology through Project No. BFM2002-02588. G.F.C. gratefully acknowledges the Spanish Ministry of Education and Science for a Juan de la Cierva grant.
\par
\hspace*{2mm}
\appendix
\section{}

For completeness, we provide here a resume of the main expressions and properties of Laguerre-Gaussian (LG) modes (see also Ref.~\onlinecite{Soskin}). Starting from Maxwell's equation in vacuum, one obtains the scalar wave equation 
\begin{eqnarray}
\nabla^{2}F = \frac{1}{c^{2}}\frac{\partial^{2} F}{\partial t^{2}} \, ,
\label{eq:SWE}
\end{eqnarray}
for any of the field components $F({\bf r},t)$ (potential vector, electric and magnetic fields, etc). The paraxial approximation assumes that if, under propagation along one given direction (e.g. the $z$-axis), the field components evolve in an {\em essentially} plane-wave fashion modulated by some slowly-varying-amplitude $u({\bf r})$, so that $F({\bf r},t)=u({\bf r})e^{ik_{0}(z-ct)}$ ($k_{0}$ is the wave vector along the $z$-axis), then, $u({\bf r})$ is a solution of the paraxial wave equation $2ik_{0}(\partial u/\partial z) + \nabla^{2}_{\perp}u=0$. Depending on the particular geometry considered, one may distinguish several families of complete, orthogonal-set of solutions which include the well-known Hermite-Gaussian beams (in Cartesian coordinates), the LG beams (in cylindrical coordinates) and the more recently discovered Ince-Gaussian beams~\cite{Bandres} (in elliptical coordinates).
\par
 LG modes satisfy the paraxial wave equation, which, in cylindrical coordinates, reads as
\begin{eqnarray}
2ik_{0}\frac{\partial u}{\partial z}+ \frac{\partial^{2}u}{\partial r^{2}}+ \frac{1}{r}\frac{\partial u}{\partial r}+ \frac{1}{r^{2}}\frac{\partial^{2}u}{\partial\phi^{2}}=0 \; .
\label{eq:PWEcyl}
\end{eqnarray}
Their whole $z$-propagating normalized profiles are
\begin{widetext}
\begin{eqnarray}
LG_{l,p}(r,\phi,z;k_{0})=\sqrt{\frac{2p!}{\pi (\vert l\vert + p)!}}\frac{1}{w(z)}\left(\frac{\sqrt{2}r}{w(z)}\right)^{\!\vert l\vert}\!L_{p}^{\vert l\vert}\!\left(\frac{2r^{2}}{w^{2}(z)}\right)\exp\left(-\frac{r^{2}}{w^{2}(z)}\right)\exp \left[il\phi+i\frac{k_{0}r^{2}}{2R(z)}+i\Phi_{G}(z)\right]\; ,
\label{eq:CARLG}
\end{eqnarray}
\end{widetext}
where $w(z)=w_{0}\sqrt{1+(z/z_{0})^{2}}$ with $w_{0}$ being the width of the mode at $z=0$, $R(z)=z\left[1+(z_{0}/z)^{2}\right]$ is the phase-front radius, $z_{0}=k_{0}w^{2}_{0}/2$ the Rayleigh range, $\Phi_{G}(z)=-(2p+\vert l\vert +1)\arctan(z/z_{0})$ the Gouy phase~\cite{Feng}, and $L_{p}^{\vert l\vert}(x)$ are the associated Laguerre polynomials
\begin{eqnarray}
L_{p}^{\vert l\vert}(x) = \sum_{m=0}^{p} (-1)^{m}\frac{(\vert l\vert + p)!}{(p-m)!\,(\vert l\vert + m)!\,m!}\,x^{m}.
\label{eq:Laguerre}
\end{eqnarray}
The indices $l=0,\pm 1,\pm 2,\ldots$ and $p=0, 1, 2,\ldots$ correspond to the winding (or topological charge) and the number of nonaxial radial nodes of the mode. The wavefront (equal phase surface) forms in space part of a helicoidal surface given by $l\phi + kz=$const. The topological charge attributed to this wavefront manifold is positive ($l>0$) for right-handed helicoids, and vice versa.
\par
Of equal importance are the normalized Fourier-transformed LG modes, which, at $z=0$, are given by
\begin{eqnarray}
\mathcal{LG}_{l,p}(\rho,\varphi)&=&\sqrt{\frac{w_{0}^{2}p!}{2\pi(\vert l\vert + p)!}}\left(\frac{w_{0}\rho}{\sqrt{2}}\right)^{\!\vert l\vert}\!L_{p}^{\vert l\vert}\!\left(\frac{w_{0}^{2}\rho^{2}}{2}\right)\nonumber\\ 
&\times & \exp\left(-\frac{w_{0}^{2}\rho^{2}}{4}\right)\exp\left[il\varphi-i\frac{\pi}{2}\left( 2p+\vert l\vert\right)\right]\; ,\nonumber\\ 
\label{eq:FOURIERLG}
\end{eqnarray}
where $\rho$ and $\varphi$ denote the frequency-space cylindrical coordinates. They fulfill the orthogonality conditions $\int d^{2}{\bf q}\,\mathcal{LG}^{*}_{l,p}({\bf q})\mathcal{LG}_{l',p'}({\bf q})=\delta_{ll'}\delta_{pp'}$. With the help of their closure relation $\sum_{l,p}\mathcal{LG}^{*}_{l,p}({\bf q})\mathcal{LG}_{l,p}({\bf q}')=\delta^{(2)}({\bf q}-{\bf q}')$ one has the following relation
\begin{widetext}
\begin{eqnarray}
e^{i\left[{\bf q}\cdot{\bf r}_{\perp}-k_{0}\vartheta^{2}z\right]} & = & 
\int d^{2}{\bf q}'\,e^{i\left[{\bf q}'\cdot{\bf r}_{\perp}-k_{0}\vartheta^{2}z\right]}\,\delta^{(2)}({\bf q}-{\bf q}') = \int d^{2}{\bf q}'\,e^{i\left[{\bf q}'\cdot{\bf r}_{\perp}-k_{0}\vartheta^{2}z\right]}\left[\sum_{l,p}\mathcal{LG}^{*}_{l,p}({\bf q})\mathcal{LG}_{l,p}({\bf q}')\right]\nonumber\\
&=&\sum_{l,p}\mathcal{LG}^{*}_{l,p}({\bf q})\! \int d^{2}{\bf q}'\,e^{i\left[{\bf q}'\cdot{\bf r}_{\perp}-k_{0}\vartheta^{2}z\right]}\mathcal{LG}_{l,p}({\bf q}')=\sum_{l,p}\mathcal{LG}^{*}_{l,p}({\bf q})\,LG_{l,p}({\bf r}_{\perp},z;k_{0}) \; ,
\nonumber
\end{eqnarray}
\end{widetext}
where in the last step we employed the propagation formula of the angular spectrum~\cite{Goodman}. This proves Eq.~(\ref{eq:CARplanetoLG}).

\end{document}